# Finding missing edges in networks based on their community structure

Bowen Yan and Steve Gregory
*Department of Computer Science, University of Bristol, Bristol BS8 1UB, England*
E-mail: yan@cs.bris.ac.uk   +44 117 954 5142

Many edge prediction methods have been proposed, based on various local or global properties of the structure of an incomplete network. Community structure is another significant feature of networks: vertices in a community are more densely connected than average. It is often true that vertices in the same community have "similar" properties, which suggests that missing edges are more likely to be found within communities than elsewhere. We use this insight to propose a strategy for edge prediction that combines existing edge prediction methods with community detection. We show that this method gives better prediction accuracy than existing edge prediction methods alone.

## I. INTRODUCTION

Many complex systems can be represented as networks, with vertices for individuals and edges denoting relations between them. Networks are analyzed in order to understand the features and properties of them, but the correctness and completeness of the dataset is often neglected. In most cases, the data are assumed to be complete and accurate. However, most real-world networks represent the results of investigations and experiments, which are often incomplete and inaccurate [1,2].

Missing edges might strongly affect the properties of networks. For example, the experimental results of protein-protein interactions have found different properties by using different methods [3]. As a result, there has been increasing interest in techniques that can find missing edges. Techniques from graph theory have been used for edge prediction; for example: methods based on vertex neighborhoods, which concern the local structure of networks, and methods based on all paths of a network, which consider the global structure of networks [4]. These methods perform well in assortative networks [4,5]. However, they only take account of one simple network structure, which may not be enough for edge prediction, and the latter methods have a high computational complexity.

Another important topological feature of networks is that vertices can be placed into communities, such that edges between vertices in the same community are dense, but are sparse between different communities [6]. It is often true that vertices in the same community have similar properties. Indeed, the main reason for the formation of community structure is assortative mixing: vertices are densely connected because they share some properties. We have previously [7] found that good community detection algorithms have error tolerance: they are unaffected by a few missing edges in a network. This suggests that the principle of community detection could be applied to edge prediction.

In this paper, we propose an approach that combines the concepts of community structure and vertex similarity. In this method, we detect the community structure of a network first. Then we use vertex similarity methods, based on local properties of the network structure, to predict missing edges separately within a community and between different communities. Throughout the paper, we restrict our attention to undirected, unweighted, and unipartite networks.

## II. RELATED WORK

Researchers' attention has increasingly focused on the evolution of networks. Networks are dynamic objects which grow and change, so many new vertices and edges appear in an original network over time. Although the study of missing edges usually focuses on static networks, both problems concentrate on vertex similarity and the structural properties of networks [8].

Many methods have already been used to predict edges in numerous fields; for example, proximity measures that are based on network topological features [4]; supervised learning methods [9]; and relational learning methods [10,11], which consider relational attributes of elements in a relational dataset. These and others are reviewed in the survey by Lü and Zhou [12].

One of the most popular concepts of proximity between two vertices, structural similarity, is based on the structure information of a network. This concerns the common features that pairs of vertices have, e.g., *Common neighbors* and *Jaccard coefficient* [4,13,14,15]; or the path information that two vertices share [16,17,18]. If two vertices do not have any common neighbors, some extended algorithms can be used to calculate their similarity [8,19], taking account of the similarity between the neighbors of two vertices.

Also, Kossinets [20] has analyzed the effect of missing data on social networks via some statistical methods. These include mean vertex degree, clustering coefficient, assortativity and average path length. Kossinets assumed that the experimental networks were complete first, and removed some vertices and edges randomly for statistical analysis.

The point that network data is not reliable has also been addressed in [21], who emphasized that it is easy to mislead the results of the experiments with incorrect data. Costenbader and Valente [22] have analyzed the stability of centrality measures on sampling networks with missing and spurious data. Borgatti *et al.* [23] have explored the robustness of centrality measures with missing data. Kim and Leskovec [24] have combined the Expectation Maximization framework with a Kronecker graphs model to predict both missing vertices and edges in a network. Other work includes research on sampling networks with missing data [25].

Finally, there are two methods, proposed by Clauset *et al.* [5] and Guimerà and Sales-Pardo [3], that have considered the concept of community structure for edge prediction. These use the maximum likelihood method to estimate sets of models with hierarchical structure or community structure to obtain the probabilities of pairs of vertices. It is not easy to apply these two methods in practice, even though they can give very good performance, because the maximum likelihood method has a high computational complexity and can only be applied to small networks.

## III. PREDICTING EDGES USING COMMUNITY STRUCTURE

Our method uses neither a complex function to calculate the probability between two arbitrary vertices nor the probabilities related to community structure. It has two steps, both of which utilize features of the network structure for edge prediction. First, the network (Fig. 1(a)) is divided into a set of communities (Fig. 1(b)), using any good community detection algorithm. Second, vertices in different communities are treated separately, using proximity measures based on vertex neighborhoods, which can calculate the probabilities for vertex pairs. We use these measures to predict edges (dashed lines in Fig. 1(c)) in the same community first, exploiting the assumption that vertices are more similar in the same community. Then we use the same measures to predict edges (dashed lines in Fig. 1(d)) between different communities, assuming these to be less likely than those in the same community. That is, we consider both the positions (within the same or different communities) and probabilities (obtained by existing vertex similarity methods) of two vertices.

For the first phase of our method, we have chosen two community detection algorithms: InfoMap [26] and OSLOM [27]. These algorithms are effective and efficient and automatically detect the "correct" number of communities. For the second phase, we have used three proximity measures (CN, AA, and RA), described in Sec. IV.C. These three methods perform well compared with other algorithms that consider the local structure of networks, and are efficient.

## IV. EXPERIMENTAL METHODOLOGY

### A. Evaluation method

We simulate "noise" (missing edges) in both artificial networks and real-world networks. This type of noise is caused by the non-response problem [3,20,21], which is often used for evaluating edge prediction methods. The general process is that an observed network is generated, from an original network, by randomly removing the required number of edges, which will be treated as "missing edges". Then, edge prediction methods are used on all vertex pairs in the observed network.

To determine the accuracy of edge prediction methods, a common measure is the AUC: the area under the ROC (receiver-operating characteristic) curve [5,28,29]. The interpretation of AUC is the probability that the score of a randomly chosen missing edge is higher than that of a randomly chosen pair of unconnected vertices. The AUC is calculated from a sequence of instances, in which each instance consists of a score and a positive or negative class label [28]. The scores are obtained by applying an edge prediction method to the vertex pairs that are not connected in the observed network. Of these unconnected vertex pairs, the "missing edges" (which were present in the original network but were removed) are classified as *true positives*, and the other vertex pairs are classified as *true negatives*.

Usually this sequence of instances is sorted by score in decreasing order: the top predicted edges are more likely to be connected. In other words, the good edge prediction methods should be able to give a high score to "missing edges", given the observed network. However, with our method, we construct two sequences, sorted separately by score. One is for the predicted intracommunity edges and the other is for the predicted edges between communities. We then concatenate the two sequences, with the predicted intracommunity edges preceding the predicted intercommunity edges. This reflects our assumption that vertex pairs within a community are more likely to be connected than those between communities.

### B. Network datasets used

First, we use the benchmark networks of Lancichinetti *et al.* [30]. These, which we shall call LFR networks, are artificial networks that are claimed to reflect the important aspects of real-world networks. The networks have several parameters. $n$ is the number of vertices; $\langle k \rangle$ and $k_{max}$ are the average and maximum degree; $\tau_1$ and $\tau_2$ are the exponents of the power-law distribution of vertex degrees and community sizes; $c_{min}$ and $c_{max}$ are the minimum and maximum community size. $\mu$ is the mixing parameter: each vertex shares a fraction $\mu$ of its edges with vertices in other communities. $o_n$ is the number of vertices that are in more than one community; $o_m$ is the number of communities that each overlapping vertex belong to.

Second, some real-world networks have been tested, listed in Table 1. Southern women [31] is a bipartite network of women and the events they attended; we use a thresholded projection of it: an edge is placed between each pair of women who attended *more than one* event together. Fb160 is a variant of the football network [6] modified to have overlapping communities: the football network is combined with a copy of itself with vertices randomly renumbered and 40% of edges randomly deleted.

**Table 1.** Real-world networks.

| Name | Ref. | Vertices | Edges |
|---|---|---|---|
| southern women | [31] | 18 | 95 |
| football | [6] | 115 | 613 |
| fb160 |  | 115 | 961 |
| netscience | [32] | 379 | 914 |
| email | [33] | 1133 | 5451 |
| blogs | [34] | 3982 | 6803 |

**C. Edge prediction methods used**

We compare our algorithm with a few edge prediction methods, three of which are used in the second phase of our algorithm. Here we define the score $\sigma$ to be the value of the relationship between two vertices. The higher the score, the more likely they are to be neighbors.

*Common neighbors* (CN). The number of common neighbors that two vertices have is a basic idea that suggests a mutual relationship between them. For example, it may be more likely that two people know each other if they have one or more acquaintances in common in a social network [35]. The function is defined as [4]:

$$\sigma_{cn} = |\Gamma_u \cap \Gamma_v|. \quad (1)$$

where $\Gamma_u$ and $\Gamma_v$ represent the set of neighbors of vertex $u$ and $v$, respectively.

*Jaccard*. This method has a similar definition, related to the probability of triangles in all connected edges of any two vertices [14], which is defined as:

$$\sigma_{jaccard} = |\Gamma_u \cap \Gamma_v| / |\Gamma_u \cup \Gamma_v|. \quad (2)$$

*Adamic and Adar* (AA). Adamic and Adar [13] proposed a similarity measure to define the similarity between two vertices in terms of the neighbors of the common neighbors of the two vertices. It is translated into [4]:

$$\sigma_{adamic} = \sum_{s \in \Gamma_u \cap \Gamma_v} \frac{1}{\log|\Gamma_s|}. \quad (3)$$

*Resource allocation index* (RA). This is a variant of the method of Adamic and Adar, which assumes that the common neighbors could transmit resources from one vertex to the other one, and is defined as [15]:

$$\sigma_{ra} = \sum_{s \in \Gamma_u \cap \Gamma_v} \frac{1}{|\Gamma_s|}. \quad (4)$$

*Preferential attachment* (PA). This shows that the probability of a connection between two arbitrary vertices is related to the number of neighbors of each vertex. The assumption of this method is that vertices prefer to connect with other vertices with a large number of neighbors. It is described as [35,36]:

$$\sigma_{pa} = |\Gamma_u||\Gamma_v|. \quad (5)$$

*Hierarchical structure method* (HRG) [5]. This method combines a maximum-likelihood method with a Markov chain Monte Carlo method to sample the hierarchical structure models with probability proportional to their likelihood from the given network. This model is a binary tree with $n$ leaves (the vertices from the given network) and $n$-1 internal nodes, and a probability $p_r$ is associated with each internal node $r$. The probability, $p_r$, of the deepest common ancestor of two vertices represents the probability of a connection between them.

Because HRG has a higher computational complexity than other methods, we can only evaluate it on some small networks in our experiments reported below.

**D. Experiments with artificial networks**

We use artificial networks to compare our method with other edge prediction methods. All results are averaged over 100 artificial networks with the same set of parameters. We set the maximum degree $k_{max}$ to $1.5\langle k\rangle$, the maximum community size $c_{max}$ to $2c_{min}$, $o_n=o_m=0$ for networks with disjoint communities, $o_n=0.4n$ and $o_m=2$ for networks with overlapping communities.

Figure 2 shows the performance of our methods, using OSLOM [27] and Infomap [26] for community detection (Infomap+CN/AA/RA, OSLOM+CN/AA/RA) compared with CN, AA, and RA on LFR networks. The results of CN, AA, and RA are similar, because they all concern the property of common neighbors only. The results of the combined methods are also similar. In contrast, it is clear that the combined methods perform much better than the basic methods. Because the basic methods are similar, we only show Infomap+AA and OSLOM+AA in the following comparisons.

Figure 3 shows a comparison of these and other edge prediction methods on LFR networks with disjoint communities. HRG is better than our algorithms when very many edges are missing in the observed networks, because then our method cannot recover the community structure. Other edge prediction methods perform worse than ours, especially PA because this method relies on the degree of each vertex; with random missing edges, it is hard to distinguish whether the vertices have a high degree or not.

Figure 4 compares our methods (Infomap+AA and OSLOM+AA) with other edge prediction methods on LFR networks with overlapping communities. Our method performs better than other methods, and slightly better than

HRG in most cases. This is because HRG is confused by the overlapping community structure, which is found more successfully by Infomap and OSLOM. Even though Infomap does not detect overlapping communities, as OSLOM does, it is very effective at finding the disjoint subsets of overlapping communities.

Figure 5 shows how the execution time varies with network size. All experiments were run on an AMD Opteron 250 at 2.4GHz. Our algorithm is much faster than HRG, which can only handle very small networks.

### E. Experiments with real-world networks

We examine the same edge prediction methods on the real-world networks listed in Table 1. Figure 6 shows the performance of our method (using Infomap) and other edge prediction methods. In the football network, HRG is slightly better than our algorithm when the fraction of edges observed is more than 0.6, but is worse than our method in the Southern women and Fb160 networks. Basically, the results of our algorithm are best in the remaining real-world networks. The only exception is that PA is slightly better than our algorithm when Email and Blogs networks have a large number of missing edges, because these two networks have a low clustering coefficient (see Table 2) and power-law degree distribution.

**Table 2.** Clustering coefficient of real-world networks.

| Network | Clustering coefficient | Average degree |
|---|---|---|
| southern women | 0.806 | 11 |
| football | 0.407 | 10.7 |
| fb160 | 0.260 | 17 |
| netscience | 0.431 | 4.8 |
| email | 0.166 | 9.6 |
| blogs | 0.102 | 3 |

### V. CONCLUSIONS

We have introduced a simple method for finding missing edges, which detects community structure first and then finds missing edges. It exploits two properties of community structure: the fact that vertices in the same community tend to be similar and the ability of good community detection algorithms to find structure despite missing edges. We have tested our method using two efficient and effective community detection algorithms for the first phase and three good vertex similarity methods for the second phase. Many modern community detection algorithms, including Infomap, are very fast, so that our method does not reduce the speed of edge prediction much (see Fig. 5). Our results, on both artificial networks and real-world networks, show that our strategy almost always improves on existing edge prediction methods. We should point out that our method relies on assortative mixing: missing edges are more likely between similar vertices, which also tend to be in the same community. In contrast, the methods of Refs. [3,5] do not make this assumption and can therefore be applied to disassortative networks.

In future, we plan to extend the technique to different types of networks: for example, weighted, directed, and multipartite networks. Edge prediction is rarely attempted in these types of network, but should be possible, using the community detection algorithms that already exist. Moreover, we may also be able to improve community detection algorithms by using edge prediction methods.

### ACKNOWLEDGEMENTS

We would like to thank the anonymous referees for their useful comments and valuable suggestions.

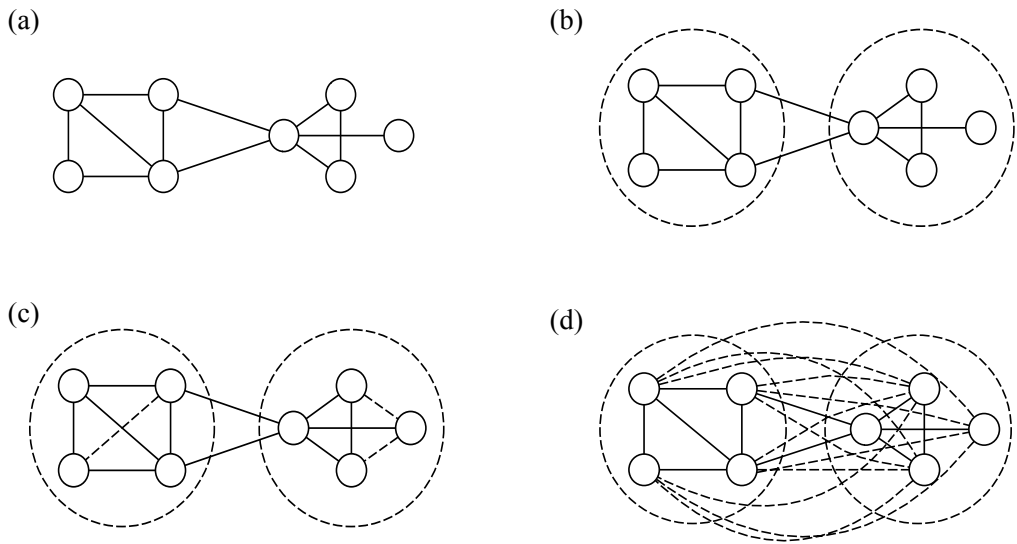

**FIG. 1.** The mechanism of the method. (a) The original network. (b) The set of communities in the network. (c) Edges need to be predicted between vertex pairs within communities (dashed lines). (d) Edges then need to be predicted between vertex pairs in different communities (dashed lines).

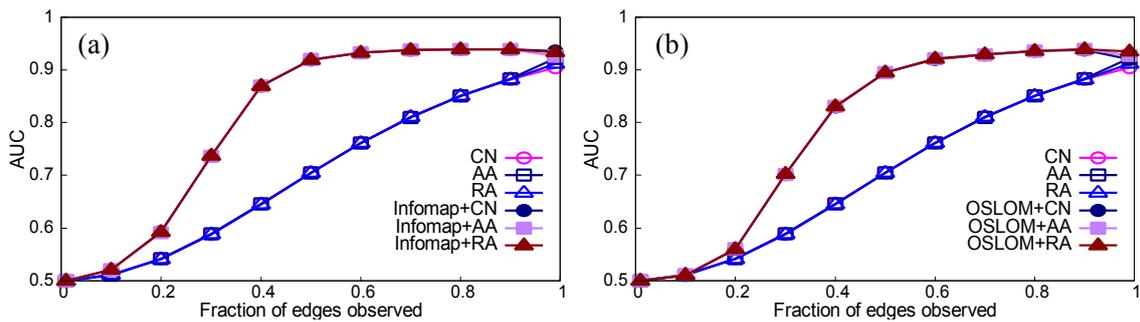

**FIG. 2.** (Color online) Performance of our algorithm (InfoMap and OSLOM with CN, AA, and RA), CN, AA, and RA on LFR networks with $n=1000$, $\langle k \rangle=10$, $k_{max}=15$, $\tau_1=2$, $\tau_2=1$, $c_{min}=20$, $c_{max}=40$, $\mu=0.1$, $o_n=0$, $o_m=0$.

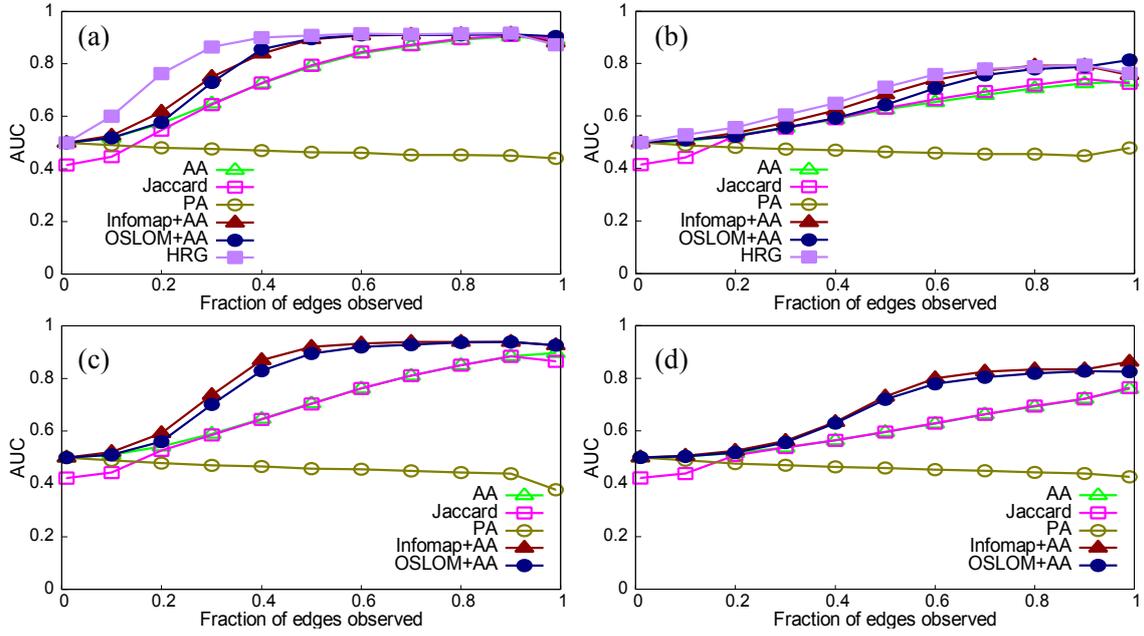

**FIG. 3.** (Color online) Performance of our algorithm (InfoMap and OSLOM with AA) and other edge prediction methods on LFR networks with disjoint communities. $\langle k \rangle=10$, $k_{max}=15$, $\tau_1=2$, $\tau_2=1$, $o_n=0$, $o_m=0$. (a) $n=100$, $c_{min}=10$, $c_{max}=20$, $\mu=0.1$. (b) $n=100$, $c_{min}=10$, $c_{max}=20$, $\mu=0.3$. (c) $n=1000$, $c_{min}=20$, $c_{max}=40$, $\mu=0.1$. (d) $n=1000$, $c_{min}=20$, $c_{max}=40$, $\mu=0.3$.

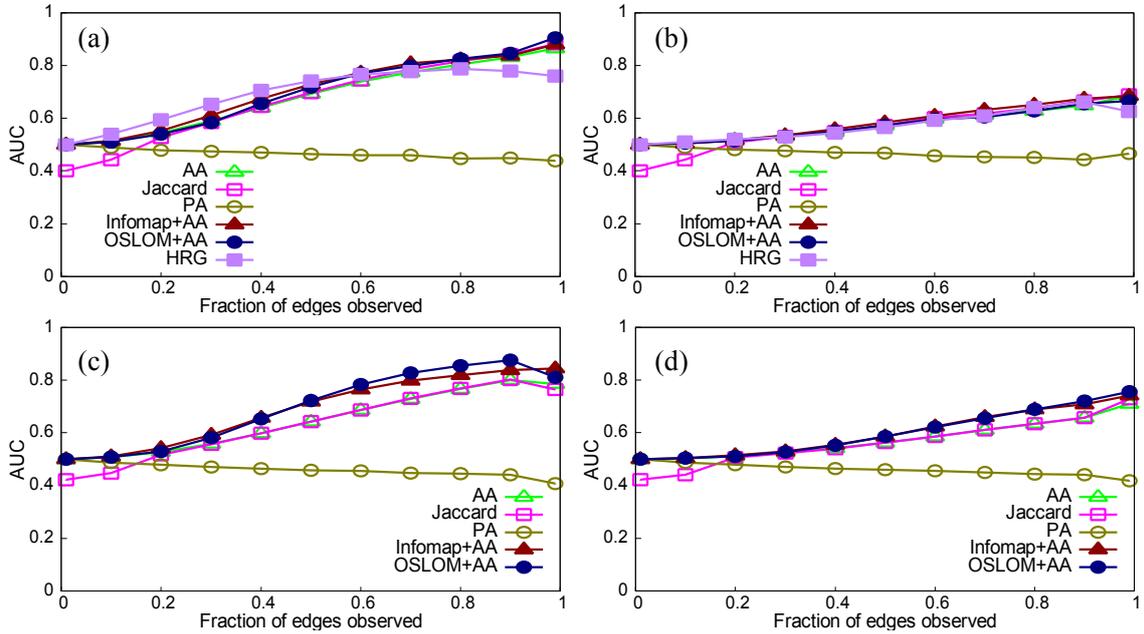

**FIG. 4.** (Color online) Performance of our algorithm (InfoMap and OSLOM with AA) and other edge prediction methods on LFR networks with overlapping communities. $\langle k \rangle=10$, $k_{max}=15$, $\tau_1=2$, $\tau_2=1$, $o_m=2$. (a) $n=100$, $c_{min}=10$, $c_{max}=20$, $\mu=0.1$, $o_n=40$. (b) $n=100$, $c_{min}=10$, $c_{max}=20$, $\mu=0.3$, $o_n=40$. (c) $n=1000$, $c_{min}=20$, $c_{max}=40$, $\mu=0.1$, $o_n=400$. (d) $n=1000$, $c_{min}=20$, $c_{max}=40$, $\mu=0.3$, $o_n=400$.

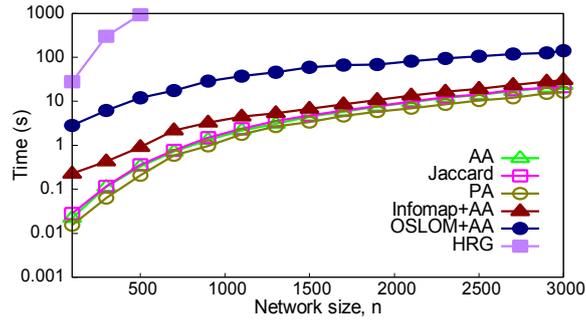

**FIG. 5.** (Color online) Execution time (seconds) of our algorithm (InfoMap and OSLOM with AA) and other edge prediction methods on LFR networks with $n=100\sim3000$, $\langle k \rangle=10$, $k_{max}=15$, $\tau_1=2$, $\tau_2=1$, $c_{min}=20$, $c_{max}=40$, $\mu=0.1$, $o_n=0$, $o_m=0$.

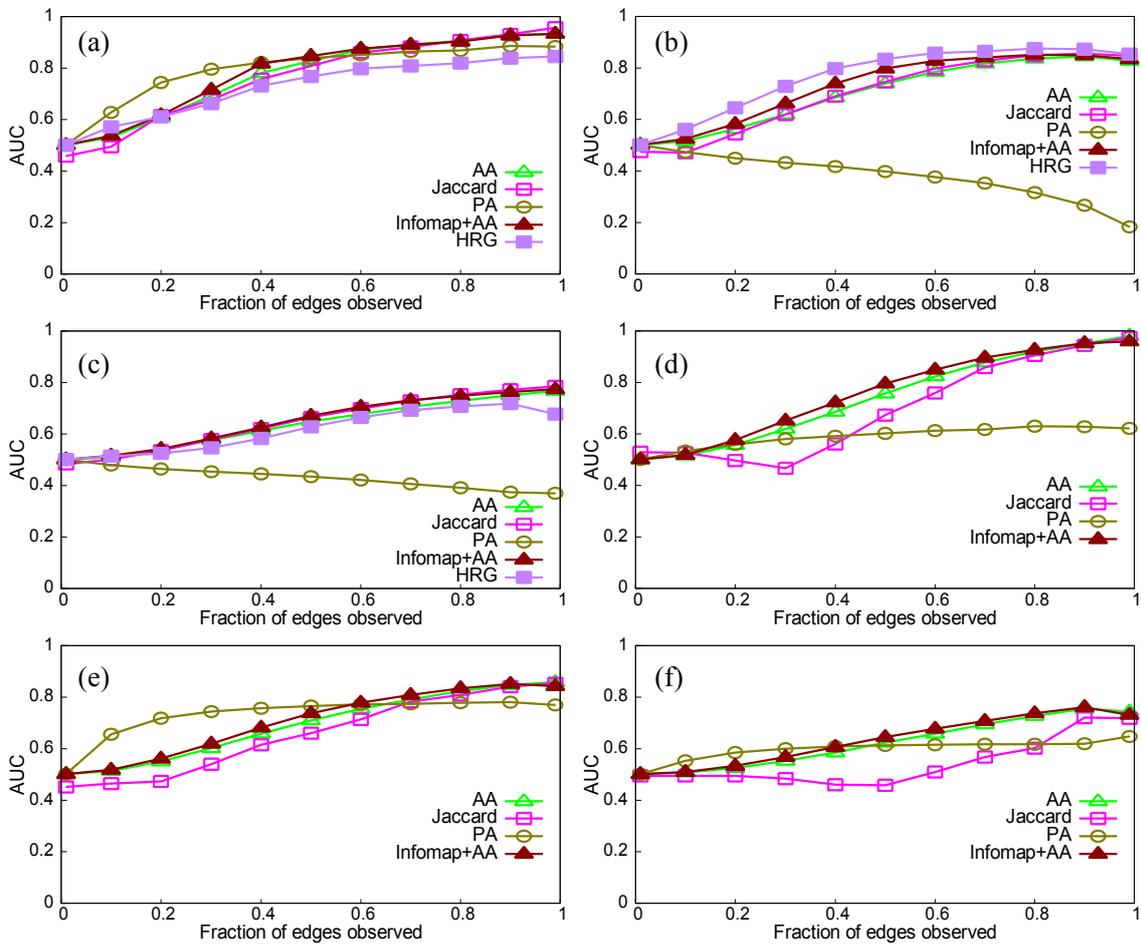

**FIG. 6** (Color online) Performance of our algorithm (InfoMap with AA) and other edge prediction methods on real-world networks. (a) Southern women. (b) Football. (c) Fb160. (d) Netscience. (e) Email. (f) Blogs.